\documentclass[sigconf]{acmart}

\AtBeginDocument{%
  }

\copyrightyear{2025}
\acmYear{2025}
\setcopyright{acmlicensed}
\acmConference[SIGIR-AP '25] {Proceedings of the 2025 Annual International ACM SIGIR Conference on Research and Development in Information Retrieval in the Asia Pacific Region}{December 7--10, 2025}{Xi'an, China.}
\acmBooktitle{Proceedings of the 2025 Annual International ACM SIGIR Conference on Research and Development in Information Retrieval in the Asia Pacific Region (SIGIR-AP '25), December 7--10, 2025, Xi'an, China}
\acmISBN{979-8-4007-2218-9/2025/12.}
\acmDOI{10.1145/3767695.3769503}
\newcommand{\IR}{IR}
\newcommand{\PL}{PL}
\newcommand{\LLM}{LLM}
\newcommand{\LLMAAJ}{LLM-as-a-Judge}
\newcommand{\ICL}{ICL}
\newcommand{\RAG}{RAG}
\newcommand{\ADT}{ADT}
\usepackage{hyperref}
\usepackage{balance}
\usepackage{pifont}
\newcommand{\xmark}{\ding{55}}
\begin{document}

\title{Which Programming Language and Model Work Best With \\ \LLMAAJ\ For Code Retrieval?}

\author{L. Roberts}
\email{rlucas7@vt.edu}
\orcid{0000-0002-5249-06842}
\affiliation{%
  \institution{Independent Researcher}
  \city{New York}
  \state{New York}
  \country{USA}
}

\author{D. Roberts}
\email{dao9853@nyu.edu}
\orcid{0000-0002-8916-6140}
\affiliation{%
  \institution{New York University}
  \city{New York}
  \state{New York}
  \country{USA}
}

\renewcommand{\shortauthors}{Roberts \& Roberts}

\begin{abstract}
Code search is an important information retrieval application.
Benefits of better code search include faster new developer on-boarding, reduced software maintenance, and ease of understanding for large repositories.
Despite improvements in search algorithms and search benchmarks, the domain of code search has lagged behind.
One reason is the high cost of human annotation for code queries and answers.
While humans may annotate search results in general text QA systems, code annotations require specialized knowledge of a programming language (\PL ), as well as domain specific software engineering knowledge.
In this work we study the use of Large Language Models (\LLM s) to retrieve code at the level of functions and to generate annotations for code search results.
We compare the impact of the retriever representation (sparse vs. semantic), programming language, and \LLM\ by comparing human annotations across several popular languages (C, Java, Javascript, Go, and Python).
We focus on repositories that implement common data structures likely to be implemented in any \PL s. 
For the same human annotations, we compare several \LLMAAJ\ models to evaluate programming language and other affinities between \LLM s.
We find that the chosen retriever and \PL\ exhibit affinities that can be leveraged to improve alignment of human and AI relevance determinations, with significant performance implications. 
We also find differences in representation (sparse vs. semantic) across \PL s that impact alignment of human and AI relevance determinations. 
We propose using transpilers to bootstrap scalable code search benchmark datasets in other \PL s and in a case study demonstrate that human-AI relevance agreement rates largely match the (worst case) human-human agreement under study.
The application code used in this work is available at
\href{https://github.com/rlucas7/code-searcher/}{this github repo}. 
\end{abstract}

\maketitle
% comment/remove this line before submission
% \listoftodos{}




\begin{teaserfigure}
  \includegraphics[scale=0.2, width=\textwidth]{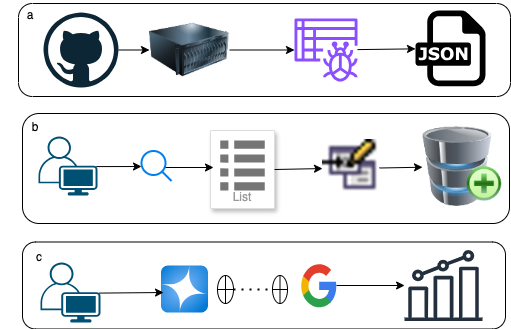}
  \caption{Code Entity Ingestion (a), Code Search and Human Annotation (b), and \IR\ Metrics Evaluation Workflows (c).}
  \Description{The 3 workflows for data ingestion, search results annotation, and metrics evaluation.}
  \label{fig:workflow}
\end{teaserfigure}

\received{1 July 2025}


\section{Introduction}
The scaling law literature \cite{fang2024scalinglawsdenseretrieval} suggests bigger benchmarks beget better code search results.
Starting from this premise we investigate \LLMAAJ\ for code search relevance determination.
Code search is a domain specific area of Information Retrieval (\IR) which aims to extract relevant code entities-in our case functions-in a corpus.
The code search domain can be separated into two main use cases: one focused on security and monitoring, typically using static analysis tools, and another aimed at human users who may be unfamiliar with the programming language or repository structure, or who prefer finding similar code examples over rewriting code from scratch.
In the static analysis setting the goal is often to identify a particular class of known security vulnerabilities in a code repository, or to notify of their introduction during new feature developments. 
In contrast, in our setting the goal is not one of monitoring but of query answering (QA).
The human information searching involves a natural or human language to formulate a query and then retrieval of indexed entities in the chosen representation, similar to other traditional information retrieval problems like open/closed question answering.
The human is assumed to be someone with some programming knowledge but may not be familiar in the particular library, code repository, or \PL .
In this work we focus on the human searching use case.
In our setting a human formulates a natural language query and enters this into an \IR\ system. 
Prior work in the code search space \cite{sadowski2015developers, stolee202510, simetal} indicates this is the predominant form of code search in developer workflows and has been so for several years. 

While QA search problems have been stimulated immensely by the support of the 
TREC series of conferences over the last decades, code search has not been a focus in the QA TREC series.
Therefore in this work we develop a test set of queries and a collection of code repositories to serve as a testbed for our research questions. 
We select repositories whose content and motivation for existence are similar across \PL s-common collections of data structures. 
A benefit of choosing these types of repositories is that the specialized knowledge for annotation is at that of an advanced undergraduate enrolled in a computer science degree program.
An additional challenge in this research is that the skill and resource limitations associated with scaling these types of benchmarks, each new \PL\ adds significant resource requirements for annotating. 
To solve the \PL\ scaling challenge we propose a means by which the benchmarks developed in one \PL\ can be leveraged to develop a scaled benchmark in an alternate \PL . 

We aim to answer the following research questions:

\begin{enumerate}
\item To what extent does the choice of \PL\ and \LLM\ for relevance annotation exhibit an affinity? If there is such an affinity which \LLM\ works best with which \PL ?
\item To what extent does the representation (sparse vs semantic) impact the ability of the \LLM\ to generate relevance that is similar to a human's relevance determination?
\item What challenges exist to scale relevance annotations across \PL s and can we scale an annotation benchmark from one \PL\ to another?
\end{enumerate}

\section{Related Work}

There are three veins of research whose prior art is relevant to this work, \IR\ \emph{annotation} approaches, (\IR ) for code, and \LLM\ based research from  the perspective of generation strategies. The latter serving as ideas for experimenting with improving \LLMAAJ\ relevance annotation performance. 

\subsection{Annotation Work}
When Amazon first released the Mechanical Turk service, TREC annotation replacement was tested \cite{alonso2009can}.
While \cite{alonso2009can} found crowd-sourcing to be a viable replacement for TREC assessment, the costs are still higher than using a programming system like an \LLM\ \cite{thomas2024largelanguagemodelsaccurately} or a multi-modal version \cite{roberts2024smart, venuto2024coderewardempoweringreinforcement} that may include modalities other than text, such as images, or audio.
A helpful survey in the \LLMAAJ\ literature is \cite{gu2024surveyllmasajudge}.
Recent work reproduced Bing's relevance assessor for open source applications \cite{upadhyay2024umbrelaumbrelaopensourcereproduction}.
Nonetheless, the claim that \LLM s obviate the need for human relevance judgments is not without detractors, for instance Clarke and Dietz \cite{clarke2024llmbasedrelevanceassessmentcant} argue there are fundamental flaws in the evaluation of relevance with \LLM s. 

Code search annotations requires specialized knowledge beyond basic computer science training, and if benchmarks are desired for purposes beyond QA systems they may require maintainability \cite{bagheri2011assessing} expertise, as well as other forms of expertise like security expertise.

\subsection{\IR\ of Code repositories, Indexing and the Query Interface}
Other aspects of \IR\ for code include indexing of the code and the way the queries interface with the indexed code. 
Two notable recent approaches are Github's indexing approach \cite{creager2023stack} and Meta's Glean tool both of which have different use cases than a human language query. 
Github's indexing approach takes a dynamic graph model and connects components as needed for querying by \emph{symbol} in the code graph-thus for github no human language query is written.
The approach of Meta's Glean tool targets an IDE environment and leverages a Query DSL called Angle \cite{glean} which then searches the indexed code for matching results. 
In contrast, Retrieval Augmentated Generation (\RAG) systems used for \LLM\ inference, have the \IR\ problem as a subservient task-often dynamically populating examples for In Context Learning (\ICL)-to improve the quality of the generated text from the \LLM .
 In \cite{wu2024repoformerselectiveretrievalrepositorylevel} the authors find that using code retrieval for (\ICL) in \RAG\ systems is sometimes detrimental for \emph{code generation} and they use a classifier to determine when to retrieve for \ICL\ in code generation tasks.

 However, to the best of the author's knowledge, no \LLMAAJ\ study has been conducted on code search relevance determinations. 
 Few Code QA studies with a public dataset exists besides the CosQA paper \cite{huang-etal-2021-cosqa}. A related work Optimizing Code Retrieval, amalgamates the few public code QA datasets \cite{li-etal-2024-optimizing} which also notes the challenges of cross \PL\ annotations and focuses instead on Python language and identifies intra and inter repository function calls as inhibiting LLM based annotations. Other approaches to improving code search include augmenting with graph structured information \cite{du2024codegrag} and dynamically choosing from varying chosen retriever \cite{tan2024prompt} during the query execution step. However, the choice of the retriever has not been studied in a controlled manner like described in this work.
 The CosQA paper is closest to our work and focuses on 19,604 natural language to python queries.
The CosQA work focuses only on Python and does not include other programming languages, but it does provide a useful, scaled code search dataset for benchmarking and testing.

\subsection{\LLM\ generated content}
Following the guidelines of \cite{van2024field} for evaluations of \LLM \ generated content, we focus on their step (i), \IR\ tasks alone.  
%Tfor human information needs and not for \RAG\ efficacy.
Therefore, evaluation of \ICL\ for code \emph{generation} is outside the scope of this manuscript.
Other works such as prompt tuning via back-propagation have been proposed recently \cite{yuksekgonul2025optimizing}. 
While many recent works have investigated the utility of \LLM s for relevance annotation in \RAG\ \cite{ni2024dirasefficientllmannotation} and search \cite{takehi2024llmassistedrelevanceassessmentsask}, to the best of the authors' knowledge, no work has investigated the use of \LLM s for relevance assessment on \emph{code search} problems. In the \LLMAAJ\ annotation workflows we leverage structured outputs \cite{liu2024weneed} to stabilize the \LLM\ generated relevance values, returning only the relevance determination for each search result. 

\section{Data Preparation}
\begin{table}
  \caption{Libraries/packages sourced from Github. The repository name column contains a link to the corresponding repository.}
  \label{tab:freq}
  \begin{tabular}{ccl}
    \toprule
    Programming Language & Repository Name & Commit-10\\
    \midrule
    C &  \href{https://github.com/srdja/Collections-C}{Collections-C} & 67a094035b \\
    Js &\href{https://github.com/montagejs/collections}{collections} & 4e19cc4890 \\
    Python & \href{https://github.com/TuTomasz/Python-Datastructures}{Python-Datastructures} & f10a879ba7 \\
    Go & \href{https://github.com/emirpasic/gods}{gods} & 8323d02ee3 \\
    Java & \href{https://github.com/lewischeng-ms/jdsl}{jdsl} & e2908c8c14 \\
  \bottomrule
\end{tabular}
\end{table}

For each of five popular programming languages (C, Javascript, Python, Go, and Java) we select and index a repository containing implementations of common data structures.
The corresponding repositories are linked in Table \ref{tab:freq} under the repository name column.
These repositories were chosen based on the programming language, similarity of implemented data structures across repositories, and open source licensing. 

We clone each of the repositories locally and index the current HEAD of the main branch.
To index the repositories we use a fork of the repository associated with \emph{The Vault} project \cite{manh2023vaultcomprehensivemultilingualdataset} which leverages the tree-sitter parser framework to generate parse trees for various \PL s. 
The purpose of \emph{The Vault} project was to clean up code in \emph{The Stack} dataset \cite{kocetkov2022stack3tbpermissively} as well as selecting high quality documentation/code data pairs for training improved \LLM s for code generation.
The Vault project found short functions and test cases did not benefit the training of \LLM s for code but for information needs these files may be answers to queries so we remove these short function filters in the indexing process.
Our fork recursively walks the repository from the root directory-where the .git file is located-and processes any files containing the specified extension for the language, for instance .go for Go, .c for the C language, etc. 
For large scale (mono-repo) repositories, extraction would need to be more sophisticated than the file extension heuristics we implemented.

Similar to other works related to code and coding agents \cite{zhang-etal-2024-codeagent}, our extracted entities are done at the function declaration level inside the respective repositories. 
While other works may study larger entities such as entire repo level structures or classes, in practice those would limit the applicability of our study to longer context models only or add additional complexity to the analysis (class vs function level).
In addition it would make cross \PL\ comparison difficult because \PL s like Go and C do not support classes and only support structures.
Therefore, we focus on indexing and search at function level only.
Each function is housed in a JSON entry and all entities are appended to a JSON Lines file for a repository.

The embeddings for each entity consist of the function and the documentation associated with the function-if any-and are stored in a database to minimize search results generation latency in the annotation process.
The semantic or sparse retriever model used to encode and retrieve the functions is the same as is used on the human language query.
The example files are included in the affiliated code artifacts.

To provide the reader with context for contents and as a prelude to subsequent relevance analyses, Table \ref{tab:adt} lists the Abstract Data Types (ADTs) provided in each repository.
The \% Docs absent row indicates the percentage of the functions over all indexed functions which do not have document strings. 
Our queries are phrased with respect to \ADT s rather than implementations, thereby providing a coarser level of abstraction, unifying the applicability of the study. In addition, this provides a means to examine synonym learning on the part of the semantic retrievers in cases where the query is phrased with one nomenclature yet the repository implements an alternate name.
An anecdote is heap vs priority queue and is described in the discussion of the results in Section \ref{sec:findings}.

The Python repository contains only a single binary search tree variation whilst the Go repository contains several variations, for example Red-Black, AVL, B-Trees, etc. 
While some \ADT s are supported across all repositories, others have only partial support, and the Trie is a rarely supported \ADT , only appearing in the Python repository.
We emphasize that the data structures may exist under one or more names but do not require modification of their API if Table \ref{tab:adt} indicates the \ADT\ is supported. 
An \ADT\ with a checkmark in Table \ref{tab:adt} indicates the repository has support for the \ADT .
%A coverage metric \cite{robertsAndGaizauskas2004} is calculated using the information from Table \ref{tab:adt}.
% NOTE: use \checkmark or \cmark (slightly different styling)
% also \xmark for x 

\begin{table}
  \caption{ADTs provided in each repository.}
  \label{tab:adt}
  \begin{tabular}{cllllll}
    \toprule
    ADT & 
    \href{https://github.com/srdja/Collections-C}{C} &
    \href{https://github.com/montagejs/collections}{Js} &
    \href{https://github.com/TuTomasz/Python-Datastructures}{Python} &
    \href{https://github.com/emirpasic/gods}{Go} & 
    \href{https://github.com/lewischeng-ms/jdsl}{Java} \\
    \midrule
     Stack &  \checkmark & \checkmark & \checkmark & \checkmark & \checkmark \\
     List &  \checkmark & \checkmark & \checkmark & \checkmark & \checkmark \\
     Set &  \checkmark & \checkmark & \xmark & \checkmark & \checkmark \\
     Map &  \checkmark & \checkmark & \xmark & \checkmark & \checkmark \\
     Ordered Set &  \checkmark & \checkmark & \xmark & \checkmark & \checkmark \\
     Tree &  \xmark & \xmark & \checkmark & \checkmark & \checkmark \\
     Queue &  \checkmark & \checkmark & \checkmark & \checkmark & \checkmark \\
     Heap &  \checkmark & \checkmark & \checkmark & \checkmark & \checkmark \\
     Trie &  \xmark & \xmark & \checkmark & \xmark & \xmark \\
     \midrule
     \% Docs Absent & 25.17 & 93.87 & 56.94 & 54.01 & 19.91 \\ 
     % \midrule
     \# Functions & 576 & 163 & 144 & 1,409 & 844 \\ 
     %\midrule
     Lines of Code & 7,285 & 1803 & 978 &  16,567 & 6,515 \\ 
     %\midrule
     \# Doc Tokens & 32,762 & 677 & 502 & 17,344 & 27,057 \\ 
     %\midrule
     \# Code Tokens & 43,257 & 12,326 & 6,724 & 132,402 & 42,482 \\ 
  \bottomrule
\end{tabular}
\end{table}

\subsection{Query Input and Human Annotations}

The query and human annotation process was conducted by the authors using the developed application framework and running the application on localhost.

\subsubsection{Queries} The sparse retrieval is done using the Huggingface bm25s package \cite{lu2024bm25sordersmagnitudefaster} and the semantic retrievers tested include Microsoft's base CodeBERT \cite{feng2020codebertpretrainedmodelprogramming} and the salesforce CodeT5+ \cite{wang2023codet5opencodelarge} models.
Other semantic retrieval models could be supported inside the application with configuration changes to the application code-e.g. model and embedding dimension-provided the semantic retrieval model is supported in the transformers library \cite{wolf-etal-2020-transformers}. All semantic retrieval methods use the cosine distance between the query and the code for ranking results. 
The re-indexing-while slow-enables any vector embedding model to be leveraged in the application as a indexer and retriever. 
For annotations we allow the human to select either relevant or not (binary).
In our metrics evaluation we select only the most recent annotation on the result chosen by the human.
In this way the human may correct an erroneous relevance judgment by re-selecting the appropriate relevance choice on any erroneously marked entity.

\subsubsection{Human Annotations} For relevance annotation determination we built a list of predefined queries to execute against the indicated repositories. Some queries can be formulated with any \ADT s, the only difference is a replacement of the name of the specific \ADT\ queried whereas others are custom for the specific \ADT .
Therefore the constructed queries span Broder's search taxonomy \cite{broder2002taxonomy} but still require domain specific expertise.
For example, a human would need to know that a stack \ADT\ could be implemented with either a linked list or a re-sizable array and then, based on the search results presented, determine if the returned entities provide sufficient information to answer the query.
The specific queries used in the relevance determinations are given in a list in the associated repository and code artifacts in the file \emph{queries.txt}.
For each query a human executes the query in the search bar and annotates all $10$ results with binary relevance values and the application stores the relevance annotations in the database locally.
We chose a cutoff of $10$ to balance simplicity, removing the need for page number considerations or below the fold impacts. 
Given that ranked results usually follow a power law distribution \cite{roberts2020expectationmaximizationframeworkyulesimon}, we expect $10$ results to cover a large portion of the relevance results for any reasonably effective retriever.
The authors are the two humans whose relevance determinations were used to generate the human relevance annotations.
The annotation guidelines included to copy and paste the query exactly into the search bar, to ensure the same listings are returned to both humans for the same queries and system configuration.
The human then chooses to label relevant or not based on the results surfaced and the human's existing knowledge and an inspection of the query results. A label of relevant is encoded as a $1$ and not-relevant as a $0$ in the database of the application.
The path to the file which contains the function as well as the function code and the documentation associated with the function are shown to the human in the result listing.
If the query has an answer in either the documentation or the source code itself, the human will select a relevant result.
If the human is not able to answer the question according to the query, then the default is a not relevant result.
In cases where queries are broadly stated, "what methods are available for a Stack?", then any function associated with the data structure is considered relevant.
Multiple distinct results may be correctly determined relevant.
In other scenarios, such as the C repository and queries related to trees, there is no possible relevant result because that particular repository does not contain the data structure. 
We note that for a larger scale study, a metric which excludes queries from the metric calculation when no result in possible \cite{terra2005comparing, robertsAndGaizauskas2004} would be import to quantify as well.
While the application stores the individual human's annotations locally, there are commands to merge relevance data if they are shared by other humans via cloud storage or other means (e.g. Bluetooth in proximity of the other human).

\subsubsection{LLM Relevances}
For machine generated annotations we adopt the \LLMAAJ\ framework described in \cite{zheng2023judgingllmasajudgemtbenchchatbot} and apply this framework to the task of search results relevance annotation. 
We used the prompt described in \cite{upadhyay2024umbrelaumbrelaopensourcereproduction} and merge all non-zero relevance values to a $1$/"relevant" value for simplicity. 
For \LLM s, we used AWS' \emph{Nova-Lite-1}, Google's \emph{Gemini-2.0-flash}, and OpenAI's \emph{GPT-4o-mini} and Meta's Llama4 models.
In preliminary testing of \LLM\ clients we noticed that some older model versions have affinities to specific prompts. 
For example, the \emph{Gemini-1.0-Pro} model would often respond with annotation results in the output format requested in the prompt used in \cite{thomas2024largelanguagemodelsaccurately} and did not follow the format requested in \cite{upadhyay2024umbrelaumbrelaopensourcereproduction} despite our use of the latter prompt.
Therefore, we used the newer \emph{Gemini-2.0-Flash} model which also showed better performance compared to older versions of the Gemini family.
API limits on the synchronous interface for some models are too strict for annotation data generation at our scale.
Therefore, we opted for batch request mode for the Google \emph{Gemini-2.0-Flash} and AWS \emph{Nova-Lite-1} models, whereas in the OpenAI \emph{GPT-4o-mini} model we did not use batch requests, despite the slower time to generate results.
Significantly, batch requests also come with cost savings. 
When possible we leverage structured outputs \cite{liu2024weneed} to stabilize the relevance annotation output format returned by the LLMs.
Once the code search queries are executed and the human annotations are captured, shell commands trigger the \LLM\ generated annotations and summary \IR\ metrics are calculated on a further shell command.
The generated relevance agreement metrics are shown in Tables \ref{tab:metrics-codebert}-\ref{tab:metrics-bm25}.

\subsection{Evaluation Metrics}

\begin{table}
  \caption{Metrics to summarize the relationship between human preference and \LLMAAJ\ relevance determinations using CodeBERT retriever.}
  \label{tab:metrics-codebert}
  \begin{tabular}{c|ccccc}
    \toprule
    Nova-lite-1  & $\kappa$ & $\tau$ / $\rho$ & RBO@10 & MAP@10\\
    \midrule
     Python & 0.052911  & 0.189055  & 0.880737  & 0.485363  \\
     C    & -0.057343  & 0.155663  & 0.929206 &  0.398335 \\
     Go   &  0.085108 &  0.226337  &  0.599856  & 0.687536 \\
     Js   &  -0.185226 & 0.018700  & 0.673431  &  0.613160 \\
     Java & -0.051613  &  0.121334  &  0.833600 & 0.657342 \\
    \midrule
    \emph{GPT-4o-mini} & $\kappa$ & $\tau$ / $\rho$ & RBO@10 & MAP@10\\
    \midrule
     Python &  -0.094620  & 0.1290278  & 0.852847 &  0.485363 \\
     C      & -0.12303  &  0.0  & 0.929107  & 0.3983350 \\
     Go     & -0.01416  &  0.1314893 & 0.658964  & 0.599856 \\
     Js     & -0.194719  & 0.0  & 0.673431  & 0.613160 \\
     Java   & -0.226857  & 0.0 & 0.832035  & 0.657342 \\
  \midrule
    \emph{Gemini-1.5} & $\kappa$ & $\tau$/$\rho$ & RBO@10 & MAP@10\\
    \midrule
     Python & 0.03735 & 0.01925 &  0.83940 & 0.21758 \\ 
     C & -0.19866 & -0.04097 &  0.89913 & 0.36550\\
     go & -0.08539 & 0.02922 &  0.67663 & 0.44928 \\
     Js & -0.30668  & -0.05405 &  0.67475 & 0.55742 \\
     Java & -0.29590 & -0.06662 &  0.74221 & 0.55722 \\
    \midrule
    \emph{Llama-4} & $\kappa$ & $\tau$/$\rho$ & RBO@10 & MAP@10\\
    \midrule
     Python &  0.02132 & 0.12156 & 0.50321 & 0.48536\\ 
     C & 0.02916 & 0.02950 & 0.67596 & 0.39834\\
     go & -0.15210 & 0.00558 & 0.42757 & 0.59986 \\
     Js & -0.06662 & 0.00185 &  0.62049 & 0.61316 \\
     Java & 0.09595 & 0.10691 & 0.69844 & 0.65734 \\
  \bottomrule

\end{tabular}
\end{table}

\begin{table}
  \caption{Metrics to summarize the relationship between human preference and \LLMAAJ\ relevance determinations using the CodeT5+ retriever.}
  \label{tab:metrics-codet5+}
  \begin{tabular}{c|ccccc}
    \toprule
    Nova-lite-1  & $\kappa$ & $\tau$ / $\rho$ & RBO@10 & MAP@10\\
    \midrule
     Python & 0.23112 & 0.33075 & 0.68241 & 0.67025 \\
     C      & 0.00075 & 0.06859 & 0.78184 & 0.49745 \\
     Go     & 0.13167 & 0.20229 & 0.60299 & 0.90146 \\
     Js     & 0.03704 & 0.11804 & 0.54745 & 0.67247 \\
     Java   & 0.08710 & 0.09525 & 0.68527 & 0.64563 \\
    \midrule
    \emph{GPT-4o-mini} & $\kappa$ & $\tau$ / $\rho$ & RBO@10 & MAP@10\\
    \midrule
     Python & -0.05479 & 0.22473 & 0.62453 & 0.67025 \\
     C      & 0.22783 & 0.24530  & 0.61883 & 0.78184 \\
     Go     & 0.00389 & 0.17256  & 0.51679 & 0.90146 \\
     Js     & -0.00464 & 0.11806 & 0.51579 & 0.67247 \\
     Java   & 0.19906 & 0.28896 & 0.69456  & 0.64562 \\
  \midrule
    \emph{Gemini-2.0} & $\kappa$ & $\tau$/$\rho$ & RBO@10 & MAP@10\\
    \midrule
     Python & -0.02140 & 0.00849 & 0.49078 & 0.67025 \\ 
     C      & 0.03108  & 0.03138 & 0.53188 & 0.78184 \\
     go     & -0.00154 & 0.03367 & 0.69062 & 0.90146 \\
     Js     & 0.03625 & 0.03698  & 0.49110 & 0.67247 \\
     Java   & -0.00504 & -0.00387 & 0.51535 & 0.64563  \\
    \midrule
    \emph{Llama-4} & $\kappa$ & $\tau$/$\rho$ & RBO@10 & MAP@10\\
    \midrule
     Python & 0.03492  & 0.11508 & 0.48657 & 0.00296 \\ 
     C      & -0.07232 & 0.01537 & 0.59313 & 0.78184 \\
     go     & 0.10634  & 0.11531 & 0.66800 & 0.90146\\
     Js     & -0.00203 & -0.00186 & 0.45948 & 0.67247 \\
     Java   & -0.27913 & -0.07869 & 0.21959 & 0.64563\\
  \bottomrule
\end{tabular}
\end{table}

\begin{table}
  \caption{Metrics to summarize the relationship between human preference and \LLMAAJ\ relevance determinations using the BM25 retriever.}
  \label{tab:metrics-bm25}
  \begin{tabular}{c|ccccc}
    \toprule
    Nova-lite-1  & $\kappa$ & $\tau$ / $\rho$ & RBO@10 & MAP@10\\
    \midrule
     Python & 0.25286 & 0.33898 & 0.88445 & 0.53101 \\
     C      & 0.07680 & 0.17046 & 0.67107 & 0.57827 \\
     Go     & 0.28438 & 0.34858 & 0.73408 & 0.71421 \\
     Js     & 0.14663 & 0.19382 & 0.86002 & 0.33028 \\
     Java   & 0.26513 & 0.34875 & 0.88454 & 0.53101 \\
    \midrule
    \emph{GPT-4o-mini} & $\kappa$ & $\tau$ / $\rho$ & RBO@10 & MAP@10\\
    \midrule
     Python & 0.02932 & 0.26277 & 0.85953 & 0.53101 \\
     C      & 0.36064 & 0.39501 & 0.77923 & 0.57827\\
     Go     & 0.10107 & 0.24199 & 0.82062 & 0.71421 \\
     Js     & 0.06904 & 0.13403 & 0.33028 & 0.88711\\
     Java   & 0.01243 & 0.24939 & 0.83319 & 0.53101 \\
  \midrule
    \emph{Gemini-2.0} & $\kappa$ & $\tau$/$\rho$ & RBO@10 & MAP@10\\
    \midrule
     Python & -0.01789 & -0.01671 & 0.50723 & 0.53101 \\ 
     C      & -0.05883 & -0.05658 & 0.55561 & 0.57827 \\
     go     &  0.00618 &  0.02326 & 0.67953 & 0.71421 \\
     Js     &  0.01178 &  0.03277 & 0.79833 & 0.33028 \\
     Java   & -0.03081 & -0.03023 & 0.51105 & 0.53101 \\
    \midrule
    \emph{Llama-4} & $\kappa$ & $\tau$/$\rho$ & RBO@10 & MAP@10\\
    \midrule
     Python & -0.05624 & 0.05443 & 0.33392 & 0.53101 \\ 
     C      & -0.17792 & -0.02671 & 0.39167 & 0.57826\\
     go     & 0.02386 & 0.12005 & 0.41517 & 0.71421 \\
     Js     & -0.02986 & 0.02002 & 0.82247 & 0.33028 \\
     Java   & -0.04586 & 0.05272 & 0.38134 & 0.53101\\
  \bottomrule
\end{tabular}
\end{table}

Metrics calculated include Cohen $\kappa$ \cite{banerjee1999beyond}, Spearman $\rho$/Kendall $\tau$ correlations, and Rank-Biased Overlap (RBO)\cite{webber2010similarity} and were chosen to conform with prior work on \LLMAAJ\ \cite{upadhyay2024umbrelaumbrelaopensourcereproduction}.  
For reproducibility, we experimented with re-executing the workflows given the human annotated inputs keeping everything else constant. The exact metrics are replicated, likely due to caching on the \LLM\ servers for the proprietary models.

Comparing Tables \ref{tab:metrics-codebert}-\ref{tab:metrics-bm25}, the CodeBERT retriever has the worst agreement over all retrievers under study with the Java \PL , as well as generally worse performance as a compared to the CodeT5+ retriever with the \LLMAAJ\ models. A notable exception to this is the Llama4 \LLMAAJ\ model which performs reasonably well with Java on the CodeBERT retriever but whose performance is outshone by the CodeT5+ retriever with the gpt-4o-mini \LLMAAJ\ model in the semantic retriever space.

\subsection{Human vs Human Relevance Annotation}

As a measure of ambiguity of relevance annotations on the query results, we compare the $2\times 2$ cross tabulations of relevance for each retriever (3) and repository (5) for a total of $15$, $2\times 2$ cross tabulated results. These values are shown in Tables \ref{tab:crossTabHumans-bm25}-\ref{tab:crossTabHumans-codebert}. 
The percentage given in the bottom left cell of each of the confusion matrices is the percentage agreement, e.g. both humans agree on relevant and both humans agree on not relevant divided by the total number of cases-the bottom right hand side entry in each confusion matrix-and rounded to the nearest hundredth. For example for the CodeBERT retriever on the C language, the first confusion matrix in Table \ref{tab:crossTabHumans-codebert} we have $(198+27) / 328 \approx 68.59756\%$ which rounds to $68.60\%$ as shown in Table \ref{tab:crossTabHumans-codebert}.

\begin{table}
  \caption{Human relevance agreements for BM25 retriever.}
  \label{tab:crossTabHumans-bm25}
  \begin{tabular}{c|cc|c}
    \toprule
    \emph{C } & not-relevant & relevant & BM25\\
    \midrule
     not-relevant & 177 & 27 & 204 \\
     relevant & 38 & 62 & 100 \\
     \midrule  
     78.61\% & 215 & 89 & 304 \\
  \bottomrule
\end{tabular}
\begin{tabular}{c|cc|c}
    \toprule
     \emph{Java} & not-relevant & relevant\\
    \midrule
     not-relevant & 215 & 12 & 227 \\
     relevant & 64 & 32 & 96 \\
     \midrule  
     76.47\% & 279 & 44 & 323 \\
  \bottomrule
\end{tabular}
\begin{tabular}{c|cc|c}
    \toprule
    \emph{js} & not-relevant & relevant \\
    \midrule
     not-relevant &   229 & 13 & 242 \\
     relevant & 33 & 12 & 45 \\
     \midrule  
    83.97\% & 262 & 25 & 287 \\
  \bottomrule
\end{tabular}
\begin{tabular}{c|cc|c}
    \toprule
    \emph{go} & not-relevant & relevant\\
    \midrule
     not-relevant & 171 & 10 & 181 \\
     relevant & 95 & 54 & 149 \\
     \midrule  
    68.18\% & 266 & 64 & 330 \\
  \bottomrule
\end{tabular}
\begin{tabular}{c|cc|c}
    \toprule
    \emph{python} & not-relevant & relevant\\
    \midrule
     not-relevant & 184 & 11 & 195 \\
     relevant & 68 & 65 & 133 \\
     \midrule  
    75.91\% & 252 & 76 & 328 \\
  \bottomrule
\end{tabular}
\end{table}

\begin{table}
  \caption{Human relevance agreements for CodeT5+ retriever.}
  \label{tab:crossTabHumans-codet5+}
  \begin{tabular}{c|cc|c}
    \toprule
    \emph{C } & not-relevant & relevant & codeT5+\\
    \midrule
     not-relevant & 124 & 35 & 159 \\
     relevant & 76 &  90 & 166 \\
     \midrule  
     65.84\% & 200 & 125 & 325 \\
  \bottomrule
\end{tabular}
\begin{tabular}{c|cc|c}
    \toprule
     \emph{Java} & not-relevant & relevant \\
    \midrule               
not-relevant & 173  & 50 & 223 \\
relevant  &  42  & 55 &  97 \\
\midrule
 68.73 \% & 215  &105 & 320 \\
  \bottomrule
\end{tabular}
\begin{tabular}{c|cc|c}
    \toprule
    \emph{js} & not-relevant & relevant \\
    \midrule
     not-relevant & 173 &  50 & 223 \\
     relevant     &  42 &  55 &  97 \\
     \midrule  
    71.25\% & 215 & 105 & 320 \\
  \bottomrule
\end{tabular}
\begin{tabular}{c|cc|c}
    \toprule
    \emph{go} & not-relevant & relevant\\
    \midrule
     not-relevant & 61 & 26 & 87 \\
     relevant & 78 & 154 & 232 \\
     \midrule  
    67.4\% &139 & 180 & 319 \\
  \bottomrule
\end{tabular}
\begin{tabular}{c|cc|c}
    \toprule
    \emph{python} & not-relevant & relevant\\
    \midrule
     not-relevant & 161 &  14 & 175 \\
     relevant     & 71 &  89 & 160 \\
     \midrule  
     74.63 \% & 232 & 103 & 335 \\                         
  \bottomrule
\end{tabular}
\end{table}

\begin{table}
  \caption{Human relevance agreements for CodeBERT retriever.}
  \label{tab:crossTabHumans-codebert}
  \begin{tabular}{c|cc|c}
    \toprule
    \emph{C } & not-relevant & relevant & CodeBERT\\
    \midrule
     not-relevant & 198 & 90 & 288 \\
     relevant & 13 & 27 & 40 \\
     \midrule  
68.6\% & 211 & 117 & 328 \\
  \bottomrule
\end{tabular}
\begin{tabular}{c|cc|c}
    \toprule
     \emph{Java} & not-relevant & relevant \\
    \midrule               
not-relevant & 121 & 129 & 250 \\
relevant  & 37 &  44 &  81 \\ 
     \midrule  
   49.85 \%  &  158 & 173 & 331 \\
  \bottomrule
\end{tabular}
\begin{tabular}{c|cc|c}
    \toprule
    \emph{js} & not-relevant & relevant \\
    \midrule
     not-relevant & 158 & 111 & 269 \\
     relevant     &   9 &  52 &  61 \\
     \midrule  
    63.63 \% & 167 & 163 & 330 \\
  \bottomrule
\end{tabular}
\begin{tabular}{c|cc|c}
    \toprule
    \emph{go} & not-relevant & relevant\\
    \midrule
     not-relevant &  193 & 61 & 254 \\
     relevant &  21 & 52 &  73 \\
     \midrule  
 74.92\% & 214 & 113 & 327 \\
  \bottomrule
\end{tabular}
\begin{tabular}{c|cc|c}
    \toprule
    \emph{python} & not-relevant & relevant\\
    \midrule
     not-relevant &  198 &  9 & 207 \\
     relevant     &   80 & 44 & 124 \\
     \midrule
 73.11\% & 278 & 53 & 331 \\
  \bottomrule
\end{tabular}
\end{table}

\section{Scaling \LLMAAJ\ across \PL s}

In this section we demonstrate a method to scale relevance annotations from a single \PL\ to another \PL . Thereby removing the need for large scale annotations by humans whose time is limited and whose knowledge of differing \PL s may vary. We leverage a transpiler to bootstrap the benchmark data in another \PL\ than python. 

\subsection{Transpilation}
\label{subsec:traspilation}
The fundamental challenge of resource limitations for cross \PL\ relevance data suggests a technological approach.
In this section we experiment with a \emph{transpiler} for scaling cross \PL\ relevance benchmark data. 
A transpiler is responsible for transforming from one \PL\ called \emph{the source} into another \PL\ called \emph{the target}.
While an \LLM\ may work as a transpiler from a human language to another human language, human to \PL\ or \emph{vice versa}, in this research we focus on leveraging the \PL\ to \PL\ transformation via a \emph{deterministic} process.
We skip CosQA records whose code cannot be transpiled directly. 

In our case we leverage the existing CosQA data, which contains 19,604 annotated human language \& python language search queries to convert the python language example into C language via a transpiler.
We then send these QA pairs-in the C language- to the \LLMAAJ\ for subsequent relevance determinations.
The relevance determinations are compared to the original relevance for the python results which are considered high quality and were verified by 3 humans manually.
The cross-tabulated values are shown in Table \ref{tab:crossTab-transpiler}. 
In the table, the rows represent the human relevance annotations from the CosQA benchmark and the columns of each confusion matrix represent the \LLMAAJ\ relevance annotations.
The vertical columns represent the \LLMAAJ\ determined relevance of the transpiled C codes.
The percentage in the bottom left is the percentage agreement between the two relevance determination methods.
For the human to human agreement metrics, the lowest agreement was with Java \PL\ and CodeBERT retriever at 49.85\%. 

The agreement metrics for the transpiled C against the original python QA pairs has a lower value of 49.53\% for Gemini-2.0-flash model and upper value 53.91\% for the nova-lite model.
Thus we find that transpiled python CosQA entries meets the lower bound of the agreement metrics between the two human annotations, albeit in different \emph{target} languages.
We note that the percentage agreement in Table \ref{tab:crossTab-transpiler} largely corresponds to the floor observed in Table \ref{tab:crossTabHumans-codebert} (Java \PL ). 

Also, given the relatively narrow range (49.54\%-53.91\%) in aggregate, on the performance of the \LLMAAJ\ approach to the transpiled C code across the various \LLMAAJ\ models, it seems reasonable to assume this is a viable starting point for code search benchmarks across programming languages regardless of the \LLMAAJ\ model chosen.

\begin{table}
  \caption{\LLMAAJ\ on transpiled CosQA data, Python transpiled to C.}
  \label{tab:crossTab-transpiler}
  \begin{tabular}{c|cc|c}
    \toprule
    \emph{nova-lite} & not-relevant & relevant\\
    \midrule
     not-relevant & 3845 & 987 & 4832 \\
     relevant     & 3171 & 1018 & 4189 \\
     \midrule
 53.91\% & 7016 & 2005 & 9021 \\
  \bottomrule
\end{tabular}
  \begin{tabular}{c|cc|c}
    \toprule
    \emph{gemini-2.0-flash} & not-relevant & relevant\\
    \midrule
     not-relevant & 2132 & 2700 & 4832 \\
     relevant     & 1784 & 2405 &4189 \\
     \midrule
  50.29\% & 3916 & 5105 & 9021 \\
  \bottomrule
\end{tabular}
  \begin{tabular}{c|cc|c}
    \toprule
    \emph{llama4} & not-relevant & relevant\\
    \midrule
     not-relevant &  1504 & 3328 & 4832 \\
     relevant     & 1225 & 2964 &4189\\
     \midrule
 49.53\% & 2729 & 6292 & 9021 \\
  \bottomrule
\end{tabular}
  \begin{tabular}{c|cc|c}
    \toprule
    \emph{gpt-4o-mini} & not-relevant & relevant\\
    \midrule
     not-relevant & 2657 & 2175 & 4832 \\
     relevant     & 2060 & 2129 &4189 \\
     \midrule
 53.05\% & 4717 & 4304 & 9021 \\
  \bottomrule
\end{tabular}
\end{table}

Some exemplars do not transpile via the $sy\_py2c$ python package and in a practical application would need either manual human conversion or perhaps LLM assistance via guided decoding, leveraging the grammar (e.g. ANSI-C formal grammar) for the target language.
Of the 19,604 examples in the CosQA data, only 9021 (46.01\%) can be transpiled via our approach, the reasons for transpilation failure often come from incompatibility of the source \PL\ language to the target \PL\ language. More detail on the statistics of the failures are quantified in Tables \ref{tab:tec} and \ref{tab:tec-detail}. The largest differences come from language differences like list comprehensions. These differences may be mitigated via rewrite rules and will be reported on elsewhere with a detailed study across transpilers.

Finally, because the code snippets in python may not have type information in C, the transpiled code has a type string (a literal) of $None$.
Therefore, improved type handling may benefit the \LLMAAJ\ for code search by providing better alignment of source and target \PL s. 
Type annotation-where absent-could be added to the CosQA dataset, or manually added to the transpiled records.  

\section{Findings and Recommendations}
\label{sec:findings}

 Similar to Szymanski et al. \cite{szymanski2024limitationsllmasajudgeapproachevaluating}, who study a (non-code) search application that also required specialized knowledge, our findings emphasize the importance of keeping humans in the loop of the evaluation process. 
 For code search, \LLM s alone may not yet provide highly aligned (to human) relevance annotations across all languages tested in this manuscript. 
 However, the relevance annotations have agreement at levels close to or slightly exceeding those of the two humans who performed the relevance annotations in some programming language. 
 In the tables, any $\kappa$ values less than $0$ indicate the agreement is \emph{worse} than random chance \cite{banerjee1999beyond}.
We now address each research question stated earlier.

\textbf{RQ1. To what extent does the choice of \PL\ and \LLM\ for relevance annotation exhibit an affinity? If there is such an affinity which \LLM\ works best with which \PL ?}

Here we find that the choice of \PL\ and \LLM\ have some affinity that varies largely based on the retriever and representation chosen. To find the best aligned \LLM\ for a given \PL\ and retriever we study Tables \ref{tab:metrics-codebert} - \ref{tab:metrics-bm25}.
For Python and Javascript, the BM25 (sparse) representation with the Nova-lite model work best for alignment of \LLMAAJ\ relevance labels.
For Go and Java, the CodeT5+ retriever (semantic) work best while the \LLMAAJ\ models that work best are Nova-lite and gpt-4o-mini respectively.
For the Go language with the CodeT5+ retriever, we note that the Llama4 open weight model is a close second best in terms of human \& AI relevance alignments.
Finally, for the C language, the BM25 (sparse) retriever with gpt-4o-mini is the best aligned with human relevance labels. 

\textbf{RQ2: To what extent does the representation (sparse vs semantic) impact the ability of the \LLM\ to generate relevance that is similar to a human's relevance determination?}

To answer this question we again turn to the results in Tables \ref{tab:metrics-codebert} - \ref{tab:metrics-bm25} and compare the best Cohen Kappa from BM25 to the best performing of the semantic retrievers under study.
The differences in terms of aligned relevance labels are given in Table \ref{tab:sparse-semantic}. 
In Table \ref{tab:sparse-semantic}, the `\% change' column indicates the percentage change in the Cohen Kappa from switching from \emph{the best sparse representation to the best semantic} representation. 

\begin{table}
  \caption{Improvement or Reduction in Human-AI Relevance Annotation Alignment, Best Sparse and Best Semantic.}
  \label{tab:sparse-semantic}
  \begin{tabular}{c|c|c|c}
    \toprule
    Language & BM25 & CodeT5+ & \% change \\
    \midrule
    Python & 0.255286 & 0.23112 & - 9.5\% \\
    C & 0.36064 & 0.22783 & -36.8 \% \\
    Go & 0.10107 & 0.13167 & 30.3 \%\\
    Js & 0.14663 & 0.03704 & -74.7 \%\\
    Java & 0.26513 & 0.19906 & -24.9 \%\\
  \bottomrule
\end{tabular}
\end{table}

Thus Go seems a viable candidate \PL\ for using semantic retrievers whereas the other \PL s may see better relevance annotation alignment by using the (sparse) BM25 retriever instead. 

\begin{figure*}[t]
    \centering
    \includegraphics[scale=0.5]{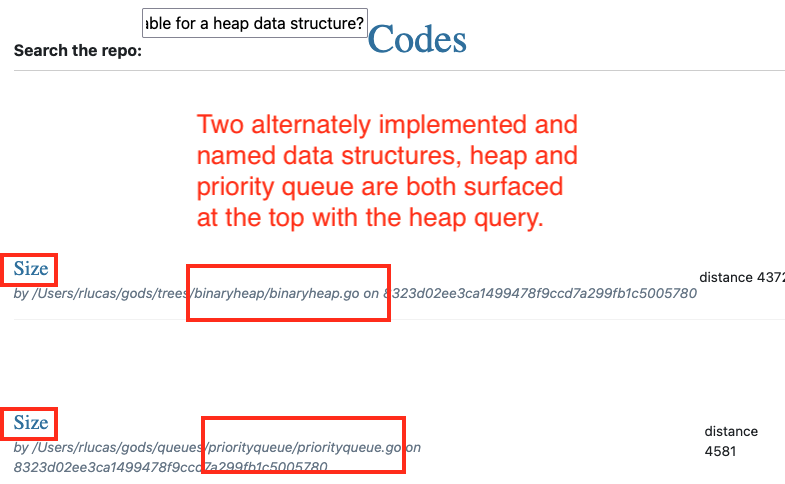}
    \caption{Semantic retriever learns the synonyms of heap and priority queue.}
    \label{fig:heap-priority-queue}
\end{figure*}

As an additional anecdote of the differences between sparse and semantic retrievers, we mention automatic synonym search. While in practice in deployed systems a configuration list may be used to handle synonyms when using BM25 as the retrieval mechanism, 
one benefit to using the semantic retrievers is that they \emph{learn the synonyms} of common pairs of terms during the training process-solving the vocabulary problem \cite{furnas1987vocabulary}. 
For example in the query, "what methods are available for a heap data structure?" where a heap is expected yet the indexed library implemented the name as $priority\_queue$ or $p\_queue$, the semantic retrievers account for this and identify the synonymous coding terms directly, without needing to resort to custom configuration lists. 
For example in the query, "what methods are available for a heap data structure?" where a heap is expected yet the indexed library implemented the name as $priority\_queue$ or $p\_queue$, the semantic retrievers identify the synonymous coding terms directly, without needing to resort to custom configuration lists. 
A screenshot with the query is shown for the programming language Go with the CodeT5+ retriever in Figure \ref{fig:heap-priority-queue}.

\textbf{RQ3: What challenges exist to scale relevance annotations across \PL s and can we scale an annotation benchmark from one \PL\ to another?}

The experiment in Section \ref{subsec:traspilation} suggests that a large relevance annotation dataset in a source \PL\ can be transpiled to a target \PL\ and the relevance determinations from \LLM s will largely perform nearly or as well as with human to human relevance agreements. 
While the initial results are promising, only 9,021 out of 19,604 QA pairs are able to be transpiled without error and they all remain with type $None$ in the \emph{target} \PL\ so that they would need type information added to compile inside a larger C program with a main function. 
The challenge remains to source a large collection of high quality code search QA pairs with relevance annotations in a language which can be deterministically transpiled to many \PL s of interest and use in the developer community. 
Also, broader support for cross language transpilation within the developer tooling community would enable the bootstrapping of cross \PL\ code search benchmark datasets.
Identification of similar repositories across \PL s remains a challenge.

Additionally, the reader may intuit that quality documentation is vital for improved relevance metrics and guess that the poor performance on Java is caused by a lack of documentation but in Table \ref{tab:adt} \% Docs Absent entries, the Java repository had the most coverage of documentation. Coverage of documentation may not be as helpful for search relevance determination as common sense suggests. 
In the repositories under study the Javascript repository has the least documentation among the extracted entities, the C repository and the Java repository have similar documentation coverage but exist at differing levels of the spectrum of \LLMAAJ\ performance and on human-human agreement.

Outside of the Go CodeT5+ case, the \emph{Gemini-2.0-flash} model performs close to random guessing on the repositories chosen.
In general we caution that the findings may vary at a larger scale along the dimensions of dataset size, \PL\, and \LLMAAJ . 

\section{Conclusion}

While it seems unlikely that automatic relevance determination will completely replace humans in determining relevance annotations in code search problems in the short term, given the costs and the potential benefits exhibited, further study is warranted.
Incremental improvements could make the \LLMAAJ \ approach viable for initial screening and removal of easy-to-judge query cases, thereby indirectly reducing costs by improving human annotation productivity.
However, as of now there is no standard way with the \LLMAAJ\ approach to separate difficult cases from easy cases with respect to any \LLMAAJ\ task. Future work to distinguish between the easy and difficult case would be helpful \LLMAAJ\ problems. 
With a reliable mechanism for distinguishing easy versus hard relevance cases, in the longer term, it seems feasible that improved \LLMAAJ \ systems could remove the majority of the human annotation tasks in this space \cite{takehi2024llmassistedrelevanceassessmentsask}.

%For code search, GTR retrievers \cite{ni2021largedualencodersgeneralizable} such as CodeT5 \cite{wang2021codet5identifierawareunifiedpretrained} and a classical, sparse method such as Okapi-BM25 \cite{robertson1994some} are both reasonable choices for a code search system.
For other applications evaluating \LLMAAJ \, claim verification \cite{olteanu2021multilingual} and recommendations \cite{dolev2025efficient} are candidates for similar evaluations. We also emphasize that various choices of information extraction are open questions for the code retrieval space. 
We chose sensible defaults of function code and associated documentation, however use of symbols in other portions of the code base, or all call sites for classes and functions would add a graph search layer similar to that implemented by \cite{creager2023stack} or the graph based embedding model in \cite{du2024codegrag}. 

A
screenshot of query results for query "what methods are available for a heap data structure?" in the go repository using CodeT5+ semantic retriever. The retriever learns the synonymous relationship in coding terms between \emph{priority queue} and
\emph{heap}.

\begin{table}
  \caption{Transpilation Exception Categories}
  \label{tab:tec}
  \begin{tabular}{lrr}
    \toprule
    Category & Frequency & \%-of Total \\
    \midrule
     Source Code$^*$ & 7418  & 70.1 \\
     Generic & 1361 & 12.9 \\
     None Not Allowed$^*$ & 842 & 8.0 \\
     Invalid Annotation$^*$ & 624 & 5.9 \\
     Attribute Error & 239 & 2.3 \\
     Syntax Error & 99 & 0.9 \\
  \bottomrule
\end{tabular}
\end{table}

\begin{table}
  \caption{Package Transpiler Detailed Exceptions.}
  \label{tab:tec-detail}
  \begin{tabular}{lrr}
    \toprule
    Node Type & Frequency & \%-of Total \\
    \midrule
ListComp &  1319 & 14.8 \\
Try      &  1058 & 11.9 \\
GeneratorExp & 911 & 10.3 \\
Constant     & 781 & 8.8 \\
With         & 624 & 7.0 \\
Slice        & 517 & 5.8 \\
Starred   & 461 & 5.2 \\
Subscript & 461 & 5.2 \\
Is        & 445 & 5.0 \\
Dict      & 402 & 4.5 \\
Raise     & 396 & 4.5 \\
In        & 325 & 3.7 \\
IsNot     & 211 & 2.4 \\
Assert    & 203 & 2.3 \\
DictComp  & 164 & 1.8 \\
Attribute & 146 & 1.6 \\
AsyncFunctionDef & 123 & 1.4 \\
Yield       & 97 & 1.1 \\
Global      & 71 & 0.8 \\
NotIn       & 64 & 0.7 \\
FunctionDef & 61 & 0.7 \\
List        & 17 & 0.2 \\
JoinedStr   & 14 & 0.2 \\
SetComp     &  4 & 0.0 \\
Set         &  3 & 0.0 \\
ClassDef    &  3 & 0.0 \\
Nonlocal    &  1 & 0.0 \\
YieldFrom   &  1 & 0.0 \\
MatMult     &  1 & 0.0 \\
\midrule
Total & 8884 & 100.0 \\
  \bottomrule
\end{tabular}
\end{table}

\newpage 
% \vspace{8cm}

%\section{Authors and Affiliations}

%Each author must be defined separately for accurate metadata
%identification.  As an exception, multiple authors may share one
%affiliation. Authors' names should not be abbreviated; use full first
%names wherever possible. Include authors' e-mail addresses whenever
%possible.

%\begin{acks}
%To Robert, for the bagels and explaining CMYK and color %spaces.
%\end{acks}
\newpage 
\bibliographystyle{ACM-Reference-Format}
\balance
\bibliography{software}

%%
%% our work had an appendix, this is now the supplementary materials-a separate file 
 
\end{document}